\documentclass[9pt,twocolumn,twoside]{opticajnl}
\journal{opticajournal} 

\setboolean{shortarticle}{true}

\usepackage{lineno}

\title{Limits and Trade-Offs of Shift-Invariant Meta-Optical Encoders for Image Compression}

\author[1]{Yubo Zhang}
\author[1]{Rui Chen}
\author[1]{Zhihao Zhou}
\author[1,2,*]{Arka Majumdar}

\affil[1]{Electrical \& Computer Engineering Department, University of Washington, Seattle, WA 98105, USA}
\affil[2]{Department of Physics, University of Washington, Seattle, WA 98105, USA}

\affil[*]{Yubo Zhang: yuboz4@uw.edu; Arka Majumdar: arka@uw.edu}

\begin{abstract}
Meta-optical encoders can reduce image data before electronic readout or transmission, but engineered point-spread functions (PSFs) do not automatically outperform conventional imaging. We study scene-agnostic, shift-invariant, linear optical encoders using both a fixed total-variation (TV) reconstruction backend and a learned YOLOv8 detection backend. Under a measurement-budget definition of compression ratio that counts all sensed samples across all channels, we compare lens imaging with spatial binning, positive random multi-channel PSFs, signed random kernels, and orthogonal multi-channel kernels. In the low-noise regime, lens-binning gives the highest reconstruction fidelity and strongest YOLOv8 detection metrics at the same compression ratio. Multi-channel encoders, however, degrade more slowly under measurement noise because the measurements are distributed across complementary channels. These results show that, for scene-agnostic incoherent imaging, engineered convolutional PSFs should be justified primarily by robustness, multiplexing, or downstream system constraints, rather than by an expectation that generic wavefront coding will outperform lens-based binning.
\end{abstract}

\setboolean{displaycopyright}{false} 

\begin{document}

\maketitle

\section{Introduction}

Optical front-ends can reduce image data before electronic readout, storage, transmission, or digital processing, making them attractive for compact and energy-constrained imaging systems \cite{optica_incoh_compression,zhang2024general,wang2024integrated,chen2023photonic}. Recent photonic and meta-optical encoders have demonstrated optical feature extraction, image compression, neural-network acceleration, and end-to-end hybrid optical design \cite{chang2018hybrid,lin2018all,huang2024photonic,wirth2025compressed,choi2025transferable,wei2024spatially,differencial_ray}. Prior end-to-end hybrid refractive-diffractive designs jointly optimize optics and reconstruction networks for task-specific computational imaging objectives, whereas we isolate passive shift-invariant convolutional encoders and compare them with lens-binning under a fixed measurement budget.  This distinction motivates a practical question: if the optical layer is limited to a passive, shift-invariant convolution, does an engineered point-spread function (PSF) improve scene-agnostic image compression relative to ordinary lens imaging followed by spatial binning?

For incoherent passive optics, the measured intensity can often be modeled as a linear mapping from object intensity to sensor intensity; when the PSF is approximately shift invariant, this mapping reduces to convolution followed by sampling. A convolutional PSF can redistribute spatial frequencies, but natural-image energy and much perceptual structure are concentrated at low spatial frequencies. Thus, for scene-agnostic reconstruction, preserving those components may be more valuable than generic measurement mixing. Conversely, multiplexed measurements can improve robustness by distributing the sensed signal across complementary channels, as in single-pixel imaging (SPI) \cite{SPI}. We study this trade-off under a fixed measurement budget and ask when convolutional meta-optical encoders are useful beyond lens-binning.

We compare four encoder families: lens imaging with spatial binning, positive random multi-channel PSFs, signed random kernels, and orthogonal multi-channel kernels. These choices span a simple diffraction-limited baseline, physically plausible non-negative multiplexed PSFs, an idealized signed-kernel comparison, and a structured multiplexed design inspired by orthogonal sampling. The reconstruction study uses a common TV-regularized linear backend \cite{tv_nju} on CelebA \cite{celeba_hq} and ARIMP4 images derived from Project Aria \cite{aria}. Because TV is not prior-free, we also test the same encoders using a learned YOLOv8 object-detection backend under a measurement-noise sweep \cite{ultralytics_yolov8}. Together, these experiments test whether the encoder ranking is tied to the optical measurement rather than only to one digital reconstruction algorithm.

\begin{figure}[ht!]
    \centering
    {\includegraphics[trim=30mm 85mm 30mm 65mm, clip, width=0.4\textwidth ]{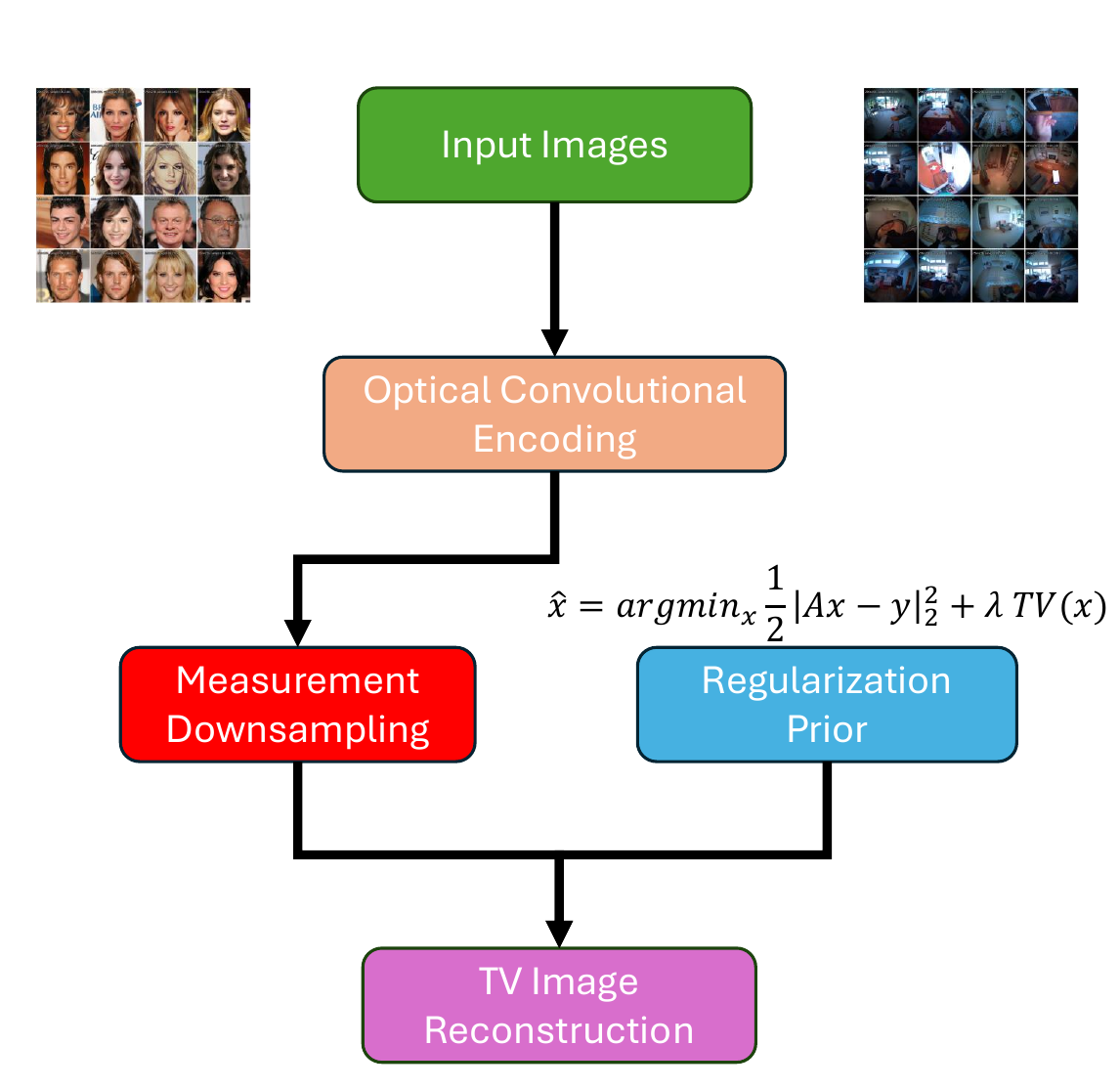}}
    \caption{Workflow for simulated optical image compression. A shift-invariant optical encoder maps the input image to compressed sensor measurements through convolutional PSFs and spatial binning. The image is then reconstructed using linear inversion with TV regularization.}
    \label{fig:1}
\end{figure}

\begin{figure}[ht!]
    \centering
    {\includegraphics[trim=35mm 25mm 35mm 25mm, clip, width=0.4\textwidth]{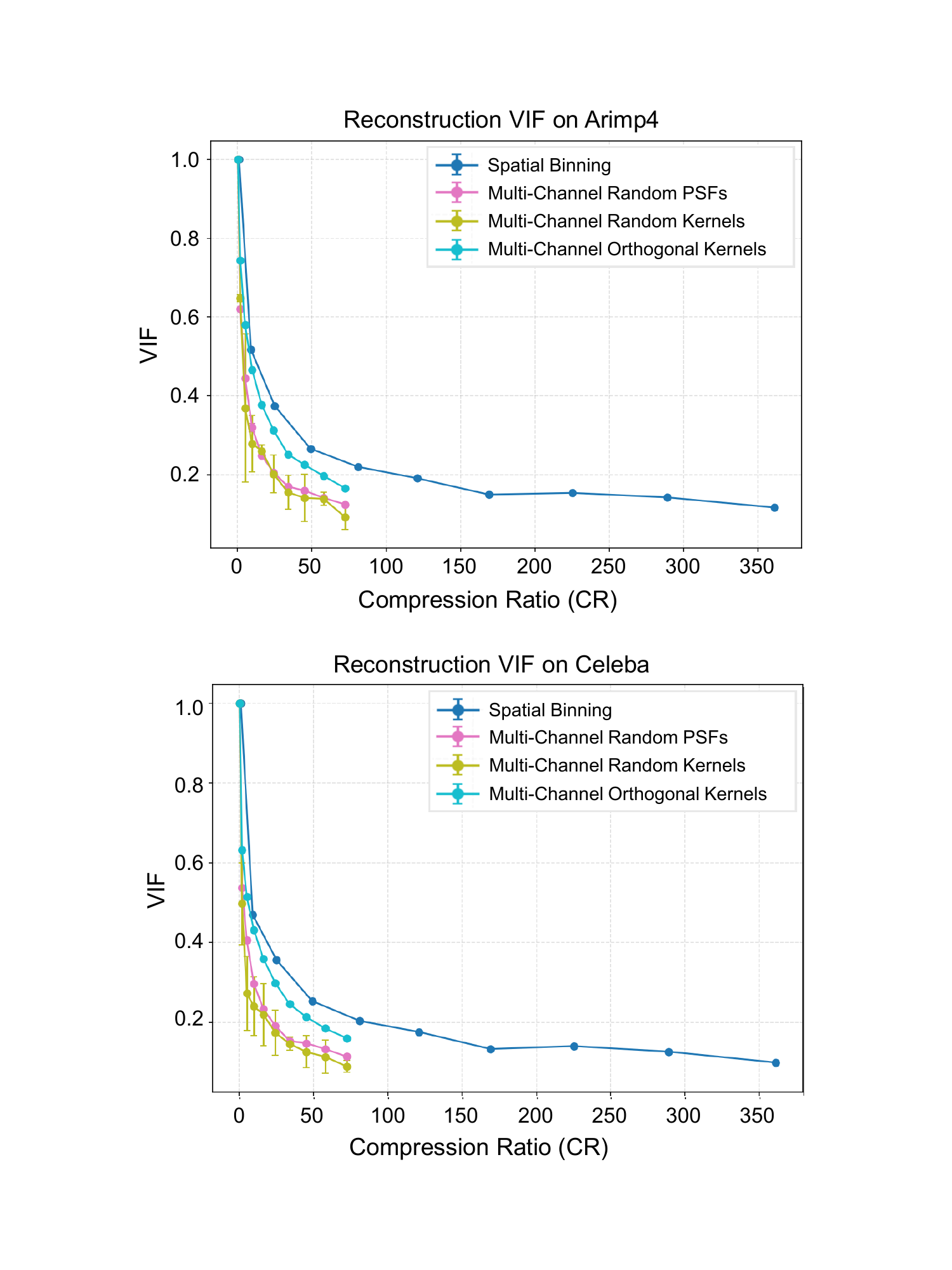}}
    \caption{Reconstruction VIF versus compression ratio on the ARIMP4 dataset (top) and CelebA dataset (bottom) for different encoder designs. The TV regularization parameter is fixed at $\lambda=10^{-1}$ for this comparison. Error bars show variation over five random seeds for random-kernel methods.}
    \label{fig:2}
\end{figure}

\section{Model and Results}

The simulated pipeline is summarized in Fig.~\ref{fig:1}. The image is encoded by one or more shift-invariant convolutional kernels and then spatially binned to reduce the number of measurements. Reconstruction solves
\begin{equation}
    \hat{x}=\mathop{\mathrm{arg\,min}}_x \|Ax-b\|_2^2 + \lambda \mathrm{TV}(x),
    \label{eq:tv}
\end{equation}
where $x$ is the reconstructed image, $b$ is the compressed measurement, and $A$ is the forward operator determined by the optical convolution and binning.

The compression ratio is defined as
\begin{equation}
    \mathrm{CR} = \frac{N_\mathrm{image}}{N_\mathrm{meas}},
\end{equation}
where $N_\mathrm{meas}$ counts all sensed samples across all channels. Thus, adding channels consumes additional detector pixels, exposures, wavelength channels, polarization states, or other multiplexed measurements. This accounting is important because a multi-channel encoder may appear better if its extra measurements are not charged against the same sensing budget. Here, the primary budget is the number of sensed measurements/readout samples; a separate fixed-total-photon-budget comparison would require implementation-specific assumptions about splitting loss, exposure time, and channel efficiency, and is outside the scope of this study. 

The four encoders are: (i) lens imaging with spatial binning; (ii) positive random multi-channel PSFs, with $n_\mathrm{channel}=5$ non-negative normalized kernels; (iii) signed random kernels, included as a mathematical comparison because passive incoherent PSFs are non-negative; and (iv) orthogonal multi-channel kernels constructed from low- to higher-spatial-frequency components. The kernel size $m$ is swept from 1 to 19. ARIMP4 and CelebA images are converted to grayscale and resized to $256\times256$ pixels; additional preprocessing and kernel-generation details are provided in the Supplement Material.

Figure~\ref{fig:2} shows visual information fidelity (VIF) \cite{vif} as a function of compression ratio for the four encoder families. The ranking is consistent across the two datasets: lens imaging followed by spatial binning gives the highest VIF in the low-noise setting, while signed random kernels are unstable and perform poorly. Because the same qualitative behavior is observed on both datasets, the remaining main-text discussion focuses on ARIMP4, with additional dataset and metric results provided in the Supplement Material.

The strong performance of lens-binning can be understood in the spatial-frequency domain. An ideal point PSF does not introduce additional pre-binning blur beyond the sampling operation itself, whereas an extended PSF attenuates or redistributes spatial frequencies according to its modulation transfer function (MTF) \cite{zhang2025applicability}. For scene-agnostic reconstruction with a generic TV prior, random or signed mixing can decorrelate measurements but does not necessarily preserve the low-frequency content that dominates reconstruction fidelity under compression.

The orthogonal multi-channel design performs better than random multi-channel encoders because its channels sample structured spatial components from low to higher frequencies. However, at the same total compression ratio, it still does not surpass lens-binning in the low-noise regime. This emphasizes the importance of counting all channels in the measurement budget: multiplexed convolutional encoders must overcome the cost of their additional measurements. In other words, decorrelation alone is not sufficient; for scene-agnostic compression, the encoder must also preserve the spatial-frequency components that the reconstruction backend can reliably use.

The TV regularization sweep is provided in Supplement Material. It reinforces that TV is a reconstruction prior: small $\lambda$ values are more sensitive to ill-conditioning and noise, while large values oversmooth the reconstruction. We use $\lambda=10^{-1}$ for the ARIMP4 results in Fig.~\ref{fig:2}.

\begin{figure}[ht!]
    \centering
    {\includegraphics[trim=30mm 55mm 35mm 65mm, clip, width=0.43\textwidth]{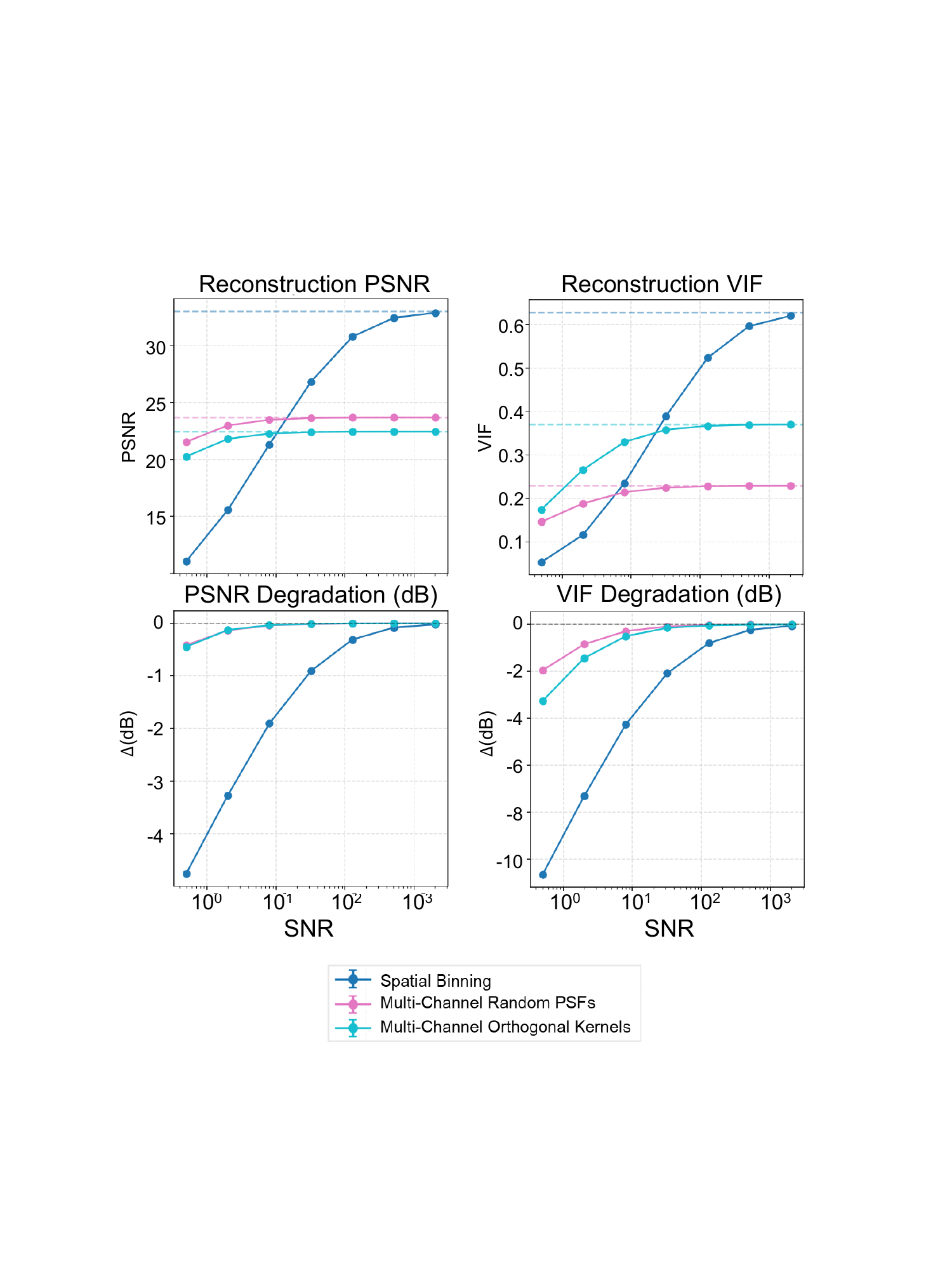}}
    \caption{Noise robustness comparison at fixed compression ratio $\mathrm{CR}=4$. The left column shows PSNR and its dB-scale loss, while the right column shows VIF and its relative dB-scale loss, computed as $10\log_{10}(\mathrm{VIF}_{\mathrm{noisy}}/\mathrm{VIF}_{\mathrm{clean}})$. Signed random kernels are omitted from this noise sweep because they are already unstable in the clean reconstruction comparisn; the noise-robustness analysis therefore focuses on the physically plausible positive PSFs and the structured orthogonal design.}
    \label{fig:4}
\end{figure}

Figure~\ref{fig:4} compares noise robustness at fixed $\mathrm{CR}=4$. At high SNR, lens-binning gives the best reconstruction quality. As SNR decreases, however, the single-channel binning strategy degrades more rapidly, while multi-channel encoders distribute the sensed signal across complementary measurements and therefore show improved robustness. This is the regime in which engineered PSFs become useful: they trade some high-SNR efficiency for reduced sensitivity to measurement perturbations.

\begin{figure}[ht!]
    \centering
    {\includegraphics[trim=32mm 22mm 35mm 30mm, clip, width=0.43\textwidth]{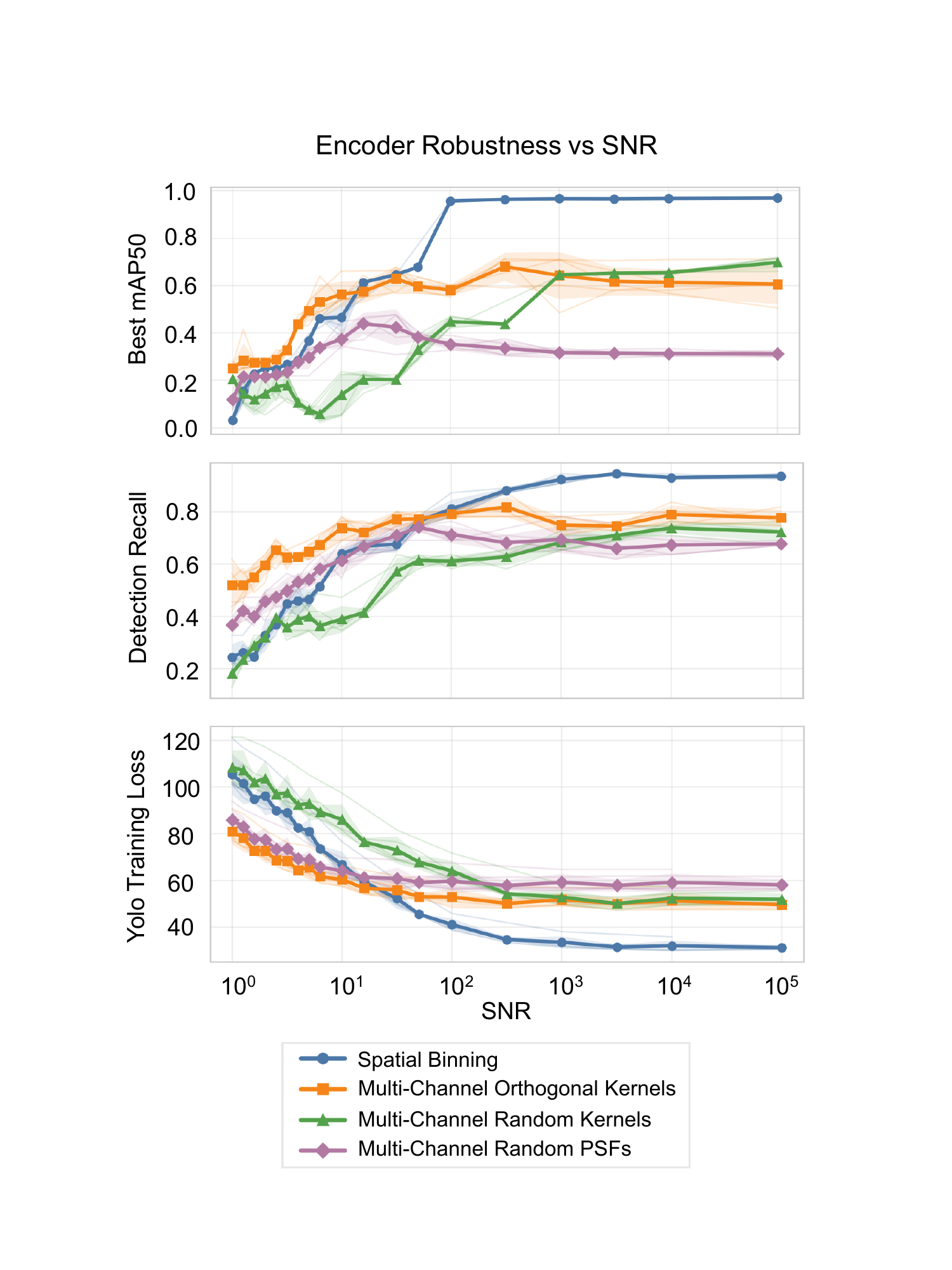}}
    \caption{Learned-backend YOLOv8 detection performance under measurement-noise sweeps for four optical encoder strategies. Top: best mAP50 as a function of SNR. Middle: detection recall. Bottom: YOLO training loss. Spatial binning gives the strongest high-SNR detection performance, while multi-channel encoders, especially orthogonal kernels, remain competitive under noisier measurement conditions. Curves summarize repeated runs over five random seeds. Shaded/faint curves represent one standard deviation.}
    \label{fig:yolo_metrics}
\end{figure}

To test whether this trade-off is specific to TV reconstruction, we also evaluate the same encoders using a learned YOLOv8 detection backend (detailed workflow in Supplement Material). Figure~\ref{fig:yolo_metrics} shows the same qualitative behavior across mAP50, recall, and training loss: spatial binning gives the best high-SNR detection performance, while multi-channel encoders, especially the orthogonal design, remain competitive at lower SNR. This small COCO128-derived experiment is a supporting robustness check, not a proof of universal generalization: it tests whether the same measurement-level trend appears after a learned detector adapts to each encoded image distribution. Thus, generic convolutional encoding does not automatically outperform lens-based binning in low-noise settings, but multi-channel measurements can provide robustness benefits under noisy acquisition.

\section{Scope and Conclusion}

The conclusions are limited to the idealized scene-agnostic, shift-invariant convolutional case. Real meta-optical systems can deviate from this model because of field-dependent PSFs over a finite field of view, chromatic aberration, fabrication errors, finite aperture, polarization dependence, and limited optical efficiency. These effects can introduce model mismatch or reduce photon throughput unless the measured spatially varying, wavelength-dependent forward operator is included in reconstruction or downstream processing.

Physical implementation also depends on kernel sign and channel structure. Positive random PSFs can be approached with passive intensity PSF engineering, while multi-channel measurements can be realized using spatially separated apertures, time, wavelength, or polarization multiplexing. Signed random kernels and any orthogonal kernels with negative components are idealized comparisons rather than direct passive incoherent PSFs; they would require differential measurements, balanced detection, bias subtraction, or another multi-measurement strategy.

Richer architectures, including spatially varying optics, cascaded diffractive systems, learned optical-digital systems, or nonlinear optoelectronic elements, may implement operators beyond a single passive convolution \cite{lin2018all,wei2024spatially,differencial_ray}; our lens-binning conclusion need not hold there. Within the present scope, however, the message is clear: under a fixed measurement budget, generic convolutional PSF engineering should not be assumed to outperform lens-based binning for scene-agnostic incoherent compression. Its most defensible role is in regimes where measurement diversity, multiplexing, or noise resilience is more valuable than preserving dominant low-frequency content with a lens-like response. This provides a practical baseline for future optical-digital co-design: before adding wavefront complexity, one should first identify the specific system constraint, such as low SNR or multiplexed acquisition, that makes the additional optical encoding worthwhile.

Code implementation of this work can be found at:
\href{https://github.com/UW-NoiseLab/Optical_Conv_Encoder.git}{GitHub repository}.

\section*{Disclosures}
The authors declare no conflicts of interest.

\section{Acknowledgement}
The research is supported by National Science Foundation (EFRI-BRAID-2223495).

\bibliography{reference}

\bibliographyfullrefs{reference}

\clearpage

\onecolumn

\clearpage

\end{document}